# Effective dimensionality of the Portevin - Le Chatelier effect

A. Chatterjee[*], N. Gayathri, A. Sarkar[1], P. Mukherjee and P. Barat

*Variable Energy Cyclotron Centre, 1/AF Bidhan Nagar, Kolkata 700064, India*

Tensile tests have been carried out by deforming polycrystalline samples of substitutional Al-2.5%Mg alloy at room temperature at a range of strain rates. The Portevin - Le Chatelier (PLC) effect was observed. From an analysis of the experimental stress versus time series data we have inferred that the dynamics of the PLC effect in a local finite time is controlled by a finite number of degrees of freedom and this effective dimension becomes reduced with increasing strain.

## 1. Introduction

The Portevin-Le Chatelier (PLC) effect was first observed at the beginning of the last century [1, 2]. Since then it has drawn huge attention due to its interesting spatio-temporal dynamics [3-8]. The phenomenon is observed in many dilute alloys. In uniaxial loading with a constant strain rate, the effect manifests itself as serrations in the stress-time (or strain) curves. This is associated with the repeated generation and propagation of plastic deformation bands. The bands mark the region of appreciable plastic deformation. In the last decade, non-local constitutive relations based on various microstructural or mechanical mechanisms leading to a spatial coupling have been used to describe this spatio-temporal effect [9-15]. However, it is now well-accepted that the microscopic origin of the PLC effect is dynamic strain aging (DSA) of materials due to the interaction between the mobile dislocations and the diffusing solute atoms. In phenomenological terms, this mechanism results in a negative strain-rate sensitivity of the flow stress. The kinetics of DSA in relation to the dislocation motion is rather well established [16-20]. However, the development and the dynamics of the localized deformation bands are less understood [21-23].

---

[*] Corresponding email address: arnomitra@veccal.ernet.in
[1] Present Address: Mechanical Metallurgy Section, Bhaba Atomic Research Centre, Mumbai – 400085, India.

Three generic types of serration can be distinguished in polycrystals, namely type A, B and C [3]. On increasing strain rates or decreasing temperature, one finds bands of type C, B, and A respectively. Type C band is randomly nucleated with large characteristic stress drops and is static in nature. Type B band has marginal spatial correlation giving the impression of hopping propagation. Finally, at higher strain rates one observes continuously propagating type A bands.

Plastic deformation of polycrystalline materials is a complex inhomogeneous process characterized by avalanches in the motion of dislocations. These types of complex dynamical systems are usually characterized by a large number of interacting components whose aggregate activity is nonlinear. These components can be identified in terms of few extrinsic and intrinsic variables. Strain, strain rate, temperature, solute concentration and specimen geometry serve as the extrinsic variables, whereas band width and band velocity, are two of the intrinsic governing factors for the dynamics of the deformation bands in the PLC effect. The nature of the involvement of these variables on the macroscopic deformation dynamics is not well established. However, it has been argued that [24] the long-range interaction of dislocations leads to highly correlated dislocation glide, which in the PLC regime manifests on the macroscopic scale in terms of bands. The collective modes associated with the propagation of the deformation bands reduce enormously the degrees of freedom of the deformation dynamics in the PLC regime. The chaotic nature of the macroscopic dynamics of the PLC effect within a definite strain rate region [10] also implies that, on this scale, the plastic deformation behaviour may be governed by a small number of relevant variables only. Again, this dynamical system exhibits self0organized criticality (SOC) with infinite degrees of freedom at higher strain rates [25]. This fact demonstrates the effect of the external variable, namely the strain rate, on the degrees of freedom of the dynamics of the PLC effect. Similarly, it may happen that the effect of some of these intrinsic variables of the deformation dynamics in the PLC regime may saturate with increasing strain and only few of them may dominate the rest of the dynamics, *i.e.* the effective dimensionality of the system may get reduced with increasing strain. Numerous techniques are available to extract the dimension of a dynamical system from the time series data [26 - 28]. Here we have adopted a statistical approach to find out the effective dimensionality of the PLC effect from the stress-time data and observe the effect of strain on it.

## 2. Experiments

In order to visualize the effective dimensionality of the PLC effect and its variation with strain we have performed uniaxial tensile tests on a Al-2.5%Mg alloy. Al-2.5% Mg sheet was prepared from an ingot by the hot-rolling process at $450^0$C followed by cold rolling and an intermediate annealing was carried out just before the final-stage cold-rolling. The rectangular sample has a gauge part $38 \times 5.15 \times 2.45$ mm$^3$ in size. Samples were tested in a servo-controlled INSTRON (model 4482) machine at four different strain rates ($1.39\times10^{-3}\sec^{-1}, 1.63\times10^{-3}\sec^{-1}, 1.80\times10^{-3}\sec^{-1}$ and $2.02\times10^{-3}\sec^{-1}$) at room temperature. These strain rates were so chosen that only type A serrations were observed during deformation in this alloy [3]. We are particularly interested in the dynamics of the type A band of the PLC effect, as it exhibits SOC which is associated with an infinite number of degrees of freedom. Thus the conventional dimensionality analysis, such as the Lyapunov spectrum [29] and the Kaplan-Yorke technique [30] cannot give any finite results for assessing the dynamics of the type A band propagation. However, it will of interest to see if one can distinguish the dominant degrees of freedom which primarily govern the dynamics of the type A deformation band. The number of these dominant degrees of freedom will be the effective dimensionality of the system. We can study the response of this effective dimensionality to the external parameter such as strain using a statistical technique: principal component analysis (PCA) [31]. Twenty-five identical samples were tested for each strain rate to generate an ensemble of data vectors. The stress time response was recorded electronically at periodic time interval of 0.05 s.

## 3. Method of Analysis

The experimental stress-time data has an increasing trend due to strain hardening. This increasing trend in the stress-strain curve has been eliminated by subtracting the average stress evolution using the method of moving average. The remaining data represents primarily the serrations present in the stress-strain curve. The discrete stress fluctuating data of a sample is divided into several segments each consisting of a hundred time points. Each segment of a hundred (*N*) time points was assigned a vector $X_i$ in our analysis. Here, we want to find out the effective dimension of the subspace spanned by the ensemble of these vectors $X_i$ arising from the testing of identical samples for a fixed range of strain using PCA. Here, the basic idea is to find the lowest dimensional subspace that in a least-squares sense optimally represents the majority of the data. Although PCA is a powerful tool for finding the dimensionality, [32] its use is limited as it is based on linear transformations. Thus, the global expansion will fail to demonstrate the intrinsic dimensionality in any nonlinear phenomenon. In such cases PCA can be applied successfully in a piecewise linear approach [33]. Since our main objective is to investigate the effect of strain on the dimensionality of the PLC effect, we have divided the entire stress-time series data set into small segments. Each segment can now be approximated as a linear one and PCA is used thereafter to find the intrinsic dimensionality.

From the fixed strain range PLC column vectors $X_i$ we constructed the covariance matrix, $\Sigma_x = \frac{1}{M}\sum_{i=1}^{M} X_i X_i^T - \bar{X}\bar{X}^T$ where $\bar{X} = \frac{1}{M}\sum_{i=1}^{M} X_i$ and M is the number of data vectors taken for the analysis. Since the covariance matrix is non-negative definite and symmetric, its eigenvalues $\lambda_i$s are non-negative and its eigenvectors $\varphi_i$s are orthogonal and span the column space of $X_i$.

## 4. Results and Discussions

It is seen that the discrete stress-strain data, obtained from different samples prepared from the same sheet of Al-2.5%Mg alloy, differ even though they were deformed at the same experimental conditions as shown in Fig. 1.

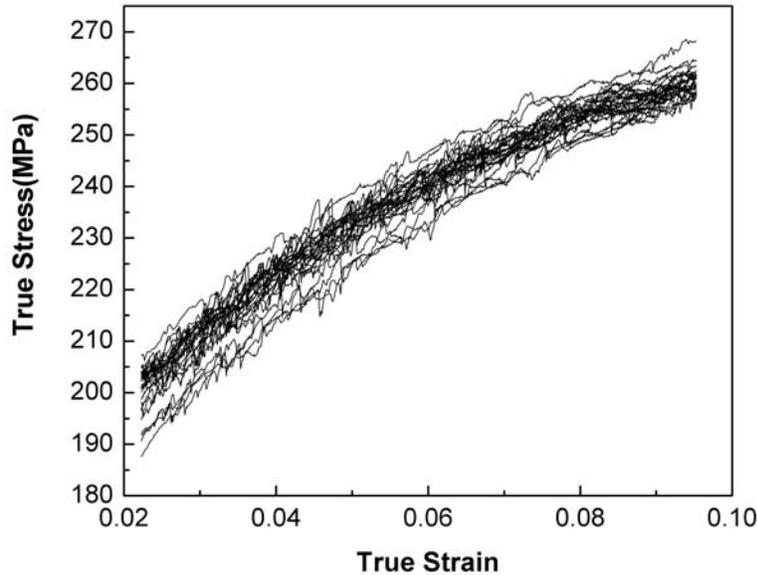

Fig.1. The segments of true stress vs. true strain curve for all the twenty-five samples deformed at a strain rate of $2.02 \times 10^{-3}$ sec$^{-1}$.

This variation arises from the spatial variation of microstructure in different samples and the complex dynamical behaviour of the type A deformation band in the PLC regime. The spatial microstructural differences among the samples were minimized by eliminating the strain hardening part from the stress-strain data. So the remaining variation in the PLC data from sample to sample is the manifestation of the complex dynamics of type A band. Before proceeding any further, some statistical analyses have been carried out on the stress-time data to investigate if the serrations are of type A deformation band. The average drop magnitudes were found to be quite small (~ 0.6 MPa) for each strain rate as is in the case of type A band. The frequency distributions of the stress drop magnitude were obtained from the stress-time data of PLC effect for all the four strain rates.

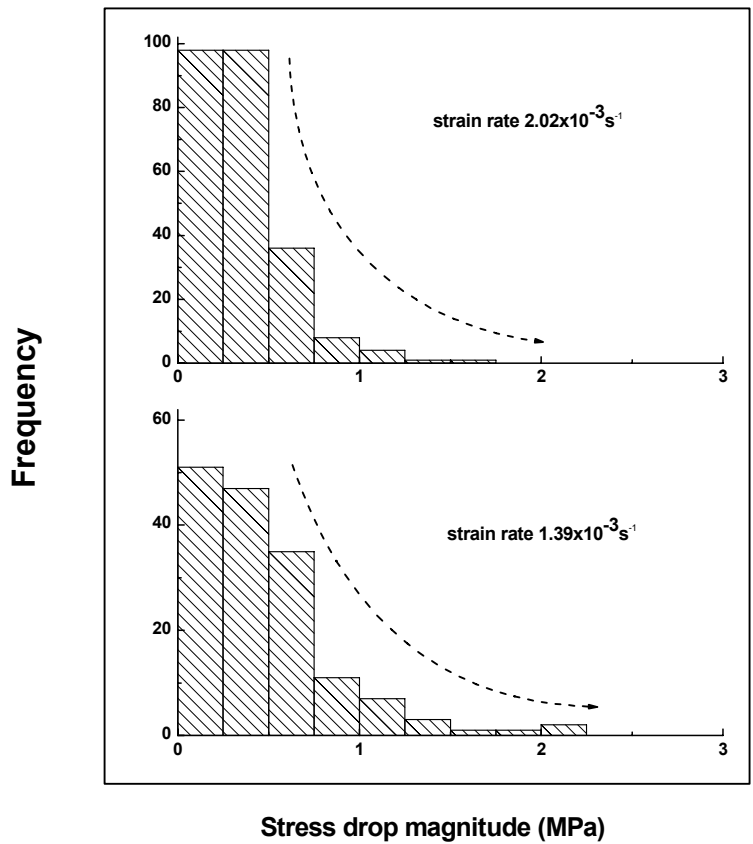

Fig.2. Frequency distribution of the stress drop magnitudes for experiments conducted at the strain rates of $1.39 \times 10^{-3} s^{-1}$ and $2.02 \times 10^{-3} s^{-1}$ on Al-2.5%Mg alloy.

Fig. 2 shows the plot of stress drop magnitude distribution for the strain rates $1.39 \times 10^{-3} s^{-1}$ and $2.02 \times 10^{-3} s^{-1}$ which are the lowest and highest limit respectively of the strain rate regime considered here. It follows that in both the cases, the distribution follows a power-law behaviour. The same nature of the distribution is obtained in all other strain rates as well. Following the formalism of Lebyodkin et al. [5,13,34], the power-law distribution of stress drop magnitude could be identified with the presence of type A band. Hence, we can definitely infer that within the strain rate regime considered here, the PLC band is type A in nature.

As the basis vectors {$\varphi_i$} are orthogonal, the components of the signal vectors $X_i$ along them are uncorrelated. Each eigenvalue $\lambda_i$ of the covariance matrix is the mean-squared projection of the $X_i$s onto the corresponding $\varphi_i$. Therefore, the spectrum {$\lambda_i$} has the information about the extent to which the signal vector explores the embedding space. The $\varphi_i$s then give the directions and the $(\lambda_i)^{\frac{1}{2}}$ the lengths of the principal axes of the ellipsoid. Thus, the effective rank of the covariance matrix defines the subspace where all the signal vectors $X_i$ arising due to the basic dynamics of the PLC effect are confined.

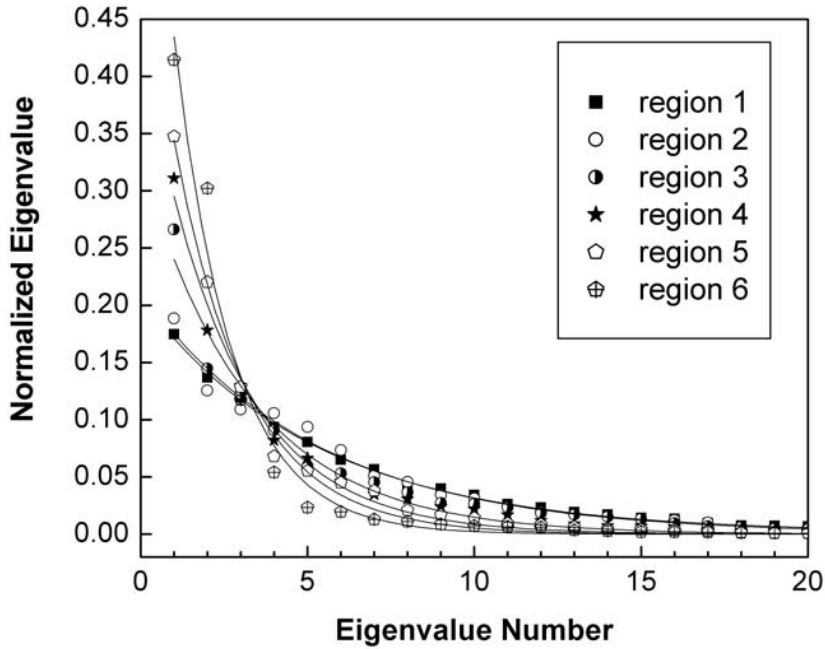

Fig.3. The normalized eigenvalue spectrum obtained from PCA for all the strain regions for the experiments conducted at the strain rate of $2.02 \times 10^{-3}$ sec$^{-1}$. The strain increases from region 1 to 6. The initial strain for the region 1 was 0.022.

Fig. 3 shows a typical plot of the normalized eigenvalue spectrum for all the strain regions for the experiment conducted at the strain rate of $2.02\times10^{-3} s^{-1}$. In the figure region 1 stands for the lowest strain region and the strain value increases from region 1 to region 6. From the figure it is clear that in the dynamics of the PLC effect some dominant eigenvalues carry a significant percent of the total variance and the normalized eigenvalue spectrum decays faster at higher strain region, i.e. the number of dominant eigenvalues decreases with increase in strain. For region 1 the largest eigenvalue represents the 18% of the total variance whereas for region 6 it is 42%. This signifies the fact that at large strain the stress fluctuations during the PLC effect become more correlated. Similar behaviours of the normalized eigenvalue spectrum are observed for the data obtained from experiments conducted at other strain rates.

By quantifying the number of eigenvectors $\boldsymbol{\varphi}_i$ necessary to represent most of the variances of $X_i$s, one can identify an effective dimension spanned by $\boldsymbol{\varphi}_i$s. For ideal noise-free system of dimension m, PCA gives m number of nonzero eigenvalues and remaining all eigenvalues are zero. Owing to the presence of noise PCA ends up giving number of nonzero singular values greater than *m*. Since they do not arise from the dynamics of the system, those values are of very small magnitude and some threshold is needed to isolate the dominant modes in PCA. Since the ordered eigenvalues $\{\lambda_i\}$ decrease rapidly with the increase in the eigenvalue number i as shown in Fig. 2, we have chosen a positive integer *D* such that:

$$D = \max \{ p: (\sum_{i=1}^{p}\lambda_i / \sum_{i=1}^{N}\lambda_i ) \le f \} \qquad (1)$$

Here $f$ is chosen as the value at which $\frac{d}{dp}(\sum_{i=1}^{p}\lambda_i / \sum_{i=1}^{N}\lambda_i) \approx 0$. The estimate of $f$ was found to be 0.95 for all the cases and that was chosen as the threshold. Thus, *D* represents the largest number of PCA modes needed to capture 95% of the total variance of the data. The quantity *D* indicates the dimension of the linear subspace that includes most of the statistical variation of the dynamics of the PLC effect. Choosing the threshold to be 95% of the variance in the data set, an estimate of the effective dimension was made with the incorporation of data vectors $X_i$ starting from 5 to 25. From this exercise we could find that the effective dimension saturates with the number of data vectors from 13 onwards, as shown in Fig. 4. This also supports us with the fact that the number of data vectors taken are sufficient to establish the dimensionality of the dynamics of the PLC effect.

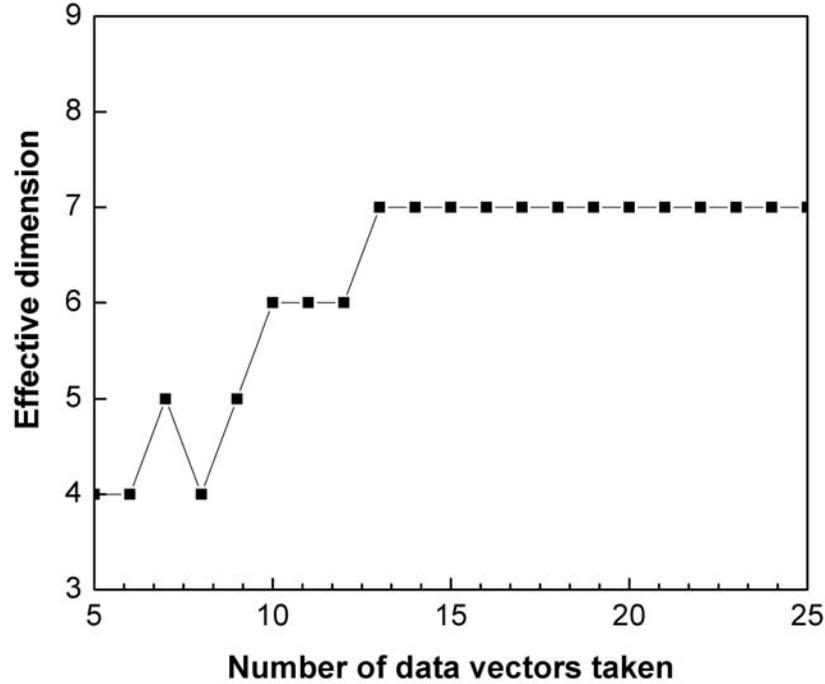

Fig.4. The saturation of the effective dimension with the number of data vectors taken for a fixed strain region for the experiments conducted at the strain rate of $2.02 \times 10^{-3} \, \text{sec}^{-1}$ when the cut off was taken as 95% of the variance of the data set.

The effective dimension so established varied from 7 to 4 for low to high strain regions for the data set taken for the four strain rates. Fig 5 represents the variation of $D$ with strain for each strain rate, where the strain value taken here represents the average strain of the respective strain region. Varying the sampling rate to 0.1 second and also changing $N$, the number of data points to be 50 for each vector $X_i$ we could get the same values of the effective dimension and its variation with strain.

    The deformation in the PLC regime is solely governed by the deformation band. Hence, several band parameters act as dynamical variables of the PLC dynamics. McCormick et.al. [35] reported a decrease in band width of type A deformation band with strain and finally observed a saturation of band width at higher strain level. Schade et al. [36] showed that the band velocity for type A band decreases with strain and reaches a plateau. Shabadi et al [8] also proved experimentally the saturation of band width and band velocity with strain. But they

mainly worked in the type B band regime. Thus, band width and band velocity do not count as the state variables of the PLC dynamics with further change in strain. These facts support the reduction of the effective degrees of freedom of the dynamics of the PLC effect with strain as identified from our analysis.

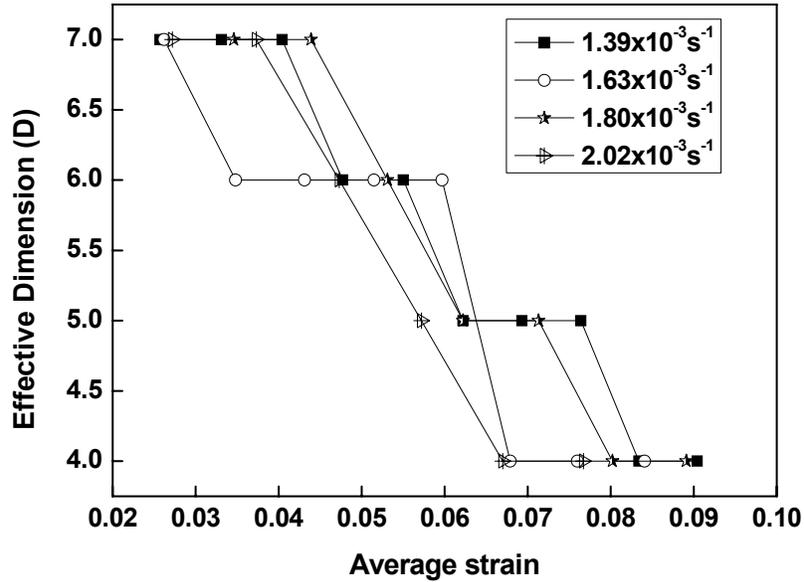

Fig.5. The variation of *D* with average strain of different strain regime for each strain rate experiment conducted on Al-Mg alloy.

It must be noted at this point that the estimated dimension is not really a dynamical dimension. It is the statistical dimension of the system.. Statistical dimension is estimated using PCA which is a linear approach. Even though we have applied PCA in piecewise linear approach to minimize the nonlinearity in the experimental stress-strain data, some marginal nonlinear coupling may be still present among the data which may decrease the degrees of freedom further. Hence, the statistical dimension provides an upper limit for the minimum number of degrees of freedom of the system.

**5. Conclusions**

In this paper we have estimated the effective dimensionality of the PLC effect in statistical sense and have demonstrated that in a local finite time only a finite number of degrees of freedom dominates this spatiotemporal dynamical system, which is designated as the 'effective dimension' of the system. This effective dimensionality gets reduced with the increase in strain. Information necessary to understand the physics of the PLC effect is not limited to the estimation of the effective dimension. Modelling the phenomenon and comparing the model results with the observed experimental data is the high road to the understanding of this complex spatio-temporal dynamical process. However, the role of the effective dimensional estimate is simply to give an idea of the number of distinct variables necessary to model this phenomenon with the best approximation.

**References: -**

1. F. Le Chatelier, Rev. de Metall. 6 (1909) p. 914.
2. A. Portevin and F. Le Chatelier, Trans. Asst. 5 (1924) p. 457.
3. K. Chihab and C. Fressengeas, Mater. Sci. Eng. A 356 (2003) p. 102.
4. M. Dablij and A. Zeghloul, Mater. Sci. Eng. A 237 (1997) p. 1.
5. M. Lebyodkin, L. Dunin-Barkowskii, Y. Brechet, Y. Estrin and L. P. Kubin, Acta Mater. 48 (2000) p. 2529.
6. A. Ziegenbein, P. Hahner and H. Neuhauser, Comput. Mater. Sci. 19 (2000) p. 27.
7. F. Chmelik, A. Ziegenbein, H. Neuhauser and P. Lukac, Mater. Sci. Eng. A 324 (2002) p. 200.
8. R Shabadi, S. Kumar, Hans J. Roven and E. S. Dwarakadasa, Mater. Sci. Eng. A 364 (2004) p. 140.
9. G. Ananthakrishna, C. Fressengeas, M. Grosbras, J. Vergnol, C. Engelke, J. Plessing, H. Neuhauser, E. Bouchaud, J. Planes and L. P. Kubin, Scr. Metall. Mater. 32 (1995) p. 1731.
10. S. Venkadesan, M. C. Valsakumar, K. P. N. Murthy and S. Rajasekar, Phys. Rev. E 54 (1996) p. 611.
11. S. J. Noronha, G. Ananthakrishna, L. Quaouire, C. Fressengeas and L. P. Kubin, Int. J. Bifurcation and Chaos 7 (1997) p. 2577.
12. G. Ananthakrishna, S. J. Noronha, C. Fressengeas and L. P. Kubin, Phys. Rev. E 60 (1999) p. 5455.
13. M. A. Lebyodkin, Y. Brechet, Y. Estrin and L. P. Kubin, Phys. Rev. Lett. 74 (1995) p. 4758.
14. G. D'Anna and F. Nori, Phys. Rev. Lett. 85 (2000) p. 4096.
15. P. Hahner, A. Ziegenbein, E. Rizzi and H. Neuhauser, Phys. Rev. B 65 (2002) p. 134109.
16. A. H. Cottrell, Dislocations and Plastic flow in Crystals, University Press, Oxford, 1953; J. Friedel, Dislocations, Pergamon Press, Oxford, 1964.
17. A. Kalk, A. Nortmann and Ch. Schwink, Philos. Mag. A 72 (1995) p. 1239.
18. A. Nortmann and Ch. Schwink, Acta Mater. 45 (1997) 2043; Acta Mater. 45 (1997) p. 2051.

19. A. Van den Beukel and U. F. Kocks, Acta Metall. 30 (1982) p. 1027.
20. J. Schlipf, Scripta Metall. Mater. 31 (1994) p. 909.
21. H. M. Zbib and E. C. Aifantis, Res. Mech. 23 (1988) 261; Scr. Metall. 22 (1988) p. 1331.
22. P. Hahner and Mater. Sci. Eng. A 164 (1993) p. 23.
23. Y. Estrin, L. P. Kubin and E. C. Aifantis, Scr. Metall. Mater. 29 (1993) p. 1147 .
24. M. Zaiser and P. Hahner, Phys Status Solidi (b) 199 (1997) p. 267.
25. M. S. Bharathi and G. Ananthakrishna, Europhys. Lett. 60 (2002) p. 234.
26. P. Grassberger and I. Procaccia, Physica D 9 (1983) p. 189.
27. J. Farmer, E. Ott and J. Yorke, Physica D 7 (1983) p. 153.
28. D. S. Broomhead, R. Jones and G.P. King, J. Phys. A: Math. Gen. 20 (1987) p. L563.
29. A. Wolf, J.B. Swift, H.L. Swinney and J.A. Vastano, Physica D 16 (1985) p. 285.
30. J. Kaplan and J. Yorke, Functional Differencial Equations and the Approximations of Fixed Points, Proceedings, Bonn, July 1978, Lecture Notes in Math. 730, H.O. Peitgen and H.O. Walther, eds., Springer, Berlin, 1978, p. 228.
31. N. Ahamed and K. R. Rao, Orthogonal Transforms for Digital Signal Processing, Springer, Berlin, 1975.
32. D. J. Patil, B. R. Hunt, E. Kalnay, J. A. Yorke and E. Ott, Phys. Rev. Lett. 86 (2001) p. 5878.
33. Keinosuke Fukunaga and David R. Olsen, IEEE transactions on Computers, c- 20 (1971) p. 2.

34. M. Lebyodkin, Y. Brechet, Y. Estrin, L.P. Kubin, Acta Mater. 44 (1996) p. 4531
35. P. G. McCormick, S. Venkadesan and C. P. Ling, Scripta Metall. Mater. 29 (1993) p 1159.
36. J. Schade, Van Westrum and A. Wijler, Acta Metall. 21 (1973) p. 1079.